\documentstyle[12pt]{article}
\def\thebibliography#1{\bigskip\section*{\centering
References\\}\bigskip\list
{\arabic{enumi}.}{\settowidth\labelwidth{#1}\leftmargin\labelwidth
\advance\leftmargin\labelsep
\usecounter{enumi}}
\def\newblock{\hskip .11em plus .33em minus .07em}
\sloppy\clubpenalty4000\widowpenalty4000
\sfcode`\.=1000\relax}

\def\op#1{\mathop{\fam0 #1}\limits}

\newcommand{\Id}{{\rm Id\,}}
\def\Ker{{\rm Ker\,}}
\newcommand{\ben}{\begin{eqnarray}}
\newcommand{\een}{\end{eqnarray}}
\newcommand{\be}{\begin{eqnarray*}}
\newcommand{\ee}{\end{eqnarray*}}
\newcommand{\bea}{\begin{eqalph}}
\newcommand{\eea}{\end{eqalph}}
\newcommand{\cL}{{\cal L}}
\newcommand{\cE}{{\cal E}}
\newcommand{\cH}{{\cal H}}
\newcommand{\cF}{{\cal F}}

\newcommand{\al}{\alpha}
\newcommand{\bt}{\beta}
\newcommand{\dl}{\delta}
\newcommand{\la}{\lambda}
\newcommand{\La}{\Lambda}

\newcommand{\om}{\omega}
\newcommand{\Om}{\Omega}
\newcommand{\m}{\mu}
\newcommand{\n}{\nu}
\newcommand{\g}{\gamma}
\newcommand{\G}{\Gamma}

\newcommand{\ve}{\varepsilon}
\newcommand{\th}{\theta}

\newcommand{\Si}{\Sigma}
\newcommand{\si}{\sigma}

\newcommand{\w}{\wedge}

\newcommand{\wt}{\widetilde}
\newcommand{\wh}{\widehat}
\newcommand{\ol}{\overline}
\newcommand{\dr}{\partial}

\newcounter{eqalph}
\newcounter{equationa}

\newenvironment{eqalph}{\stepcounter{equation}
\setcounter{equationa}{\value{equation}}
\setcounter{equation}{0}

\begin{eqnarray}}{\end{eqnarray}
\setcounter{equation}{\value{equationa}}}

\begin{document}
\hbox{}

\centerline{\bf\large ENERGY-MOMENTUM CONSERVATION LAWS}
\medskip

\centerline{\bf\large IN AFFINE-METRIC GRAVITATION THEORY.}
\bigskip

\centerline{\bf Gennadi A Sardanashvily}
\medskip

\centerline{Department of Theoretical Physics, Physics Faculty,}

\centerline{Moscow State University, 117234 Moscow, Russia}

\centerline{E-mail: sard@grav.phys.msu.su}
\bigskip

\begin{abstract} \noindent
The Lagrangian formulation of field theory does not provide any universal
energy-momentum conservation law in order to analize that in gravitation
theory. In Lagrangian field theory, we get different identities involving
different stress energy-momentum tensors which however are not conserved,
otherwise in the covariant multimomentum Hamiltonian formalism.
In the framework of this formalism, we have the fundamental identity whose
restriction to a constraint space can be treated the energy-momentum
transformation law. This identity remains true also for gravity. Thus, the
tools are at hand to investigate the energy-momentum conservation laws
in gravitation theory. The key point consists in the feature of a metric
gravitational field whose canonical momenta on the constraint space are equal
to zero.
\end{abstract}

\section{Introduction}

In Hamiltonian mechanics, there is the conventional energy transformation
law
\begin{equation}
\frac{d\cH}{dt}=\frac{\dr\cH}{\dr t} \label{E1}
\end{equation}
on solutions of the Hamilton equations, otherwise in field theory.

The standard Hamiltonian formalism has been applied to field theory. In the
straightforward manner, it takes the form of the instantaneous
Hamiltonian formalism when canonical variables are field functions at a
given instant of time. The corresponding phase space is
infinite-dimensional, so that the Hamilton equations in the bracket form are
not the familiar differential equations, adequate to the Euler-Lagrange
field equations.

In Lagrangian field theory, we have no conventional energy-momentum
transformation law. One gets different identities which involve
different stress energy-momentum tensors, in particular, different canonical
energy-momentum tensors.  Moreover, one can not say {\it a priori} what
is really concerved.

We follow the generally accepted geometric description of classical
fields by sections of fibred manifolds $Y\to X.$
Their dynamics is phrased in terms of jet spaces \cite{got,kol,sard,8sar,sau}.
Given a fibred manifold
$Y\to X$, the $k$-order jet space $J^kY$ of $Y$
comprises the equivalence classes
$j^k_xs$, $x\in X$, of sections $s$ of $Y$ identified by the first $(k+1)$
terms of their Taylor series at a point $x$.
It is a finite-dimensional smooth manifold. Recall that
a $k$-order differential operator on sections of a fibred manifold $Y$,
by definition, is a morphism of $J^kY$ to a vector bundle over $X$.
As a consequence, the dynamics of field systems is played out
on finite-dimensional configuration and phase spaces.

In field theory, we can restrict ourselves
to the first order  Lagrangian formalism when the configuration space
is $J^1Y$. Given fibred coordinates $(x^\m, y^i)$
of $Y$, the jet space $J^1Y$ is endowed with the adapted  coordinates
$ (x^\m, y^i, y^i_\m)$:
\[{y'}^i_\la=(\frac{\dr {y'}^i}{\dr y^j}y_\m^j +
\frac{\dr{y'}^i}{\dr x^\m})\frac{\dr x^\m}{\dr{x'}^\la}.\]
A first order Lagrangian density on the configuration space $J^1Y$ is
represented by a horizontal exterior density
\[L=\cL(x^\m, y^i, y^i_\m)\om, \qquad \om=dx^1\w ...\w dx^n,
\qquad n=\dim X.\]
The corresponding first order Euler-Lagrange equations for sections
$\ol s$ of the fibred jet manifold $J^1Y\to X$ read
\ben &&\dr_\la\ol s^i=\ol s^i_\la, \nonumber\\
&& \dr_i\cL-(\dr_\la+\ol s^j_\la\dr_j
+\dr_\la\ol s^j_\m\dr^\m_j)\dr^\la_i\cL=0. \label{306}\een

We consider the
Lie derivatives of Lagrangian densities in order to obtain differential
conservation laws. Let
\[ u=u^\m\dr_\m + u^i\dr_i\]
be a vector field on a fibred manifold $Y$ and $\ol u$ its jet lift
(\ref{1.21})
onto the fibred jet manifold $J^1Y\to X$. Given a
Lagrangian density $L$ on $J^1Y$, let us computer the Lie derivative
${\bf L}_{\ol u}L$. On solutions $\ol s$ of the first order Euler-Lagrange
equations (\ref{306}), we have the equality
\begin{equation}
\ol s^*{\bf L}_{\ol u}L= \frac{d}{dx^\la}[\pi^\la_i(\ol s)(u^i-u^\m \ol
s^i_\m) +u^\la\cL (\ol s)]\om, \qquad \pi^\m_i=\dr^\m_i\cL. \label{502}
\end{equation}

In particular, if $u$ is a vertical vector field such that
\[{\bf L}_{\ol u}L=0,\]
the equality (\ref{502}) takes the form of the current
conservation law
\begin{equation}
\frac{d}{dx^\la}[u^i\pi^\la_i(\ol s)]=0. \label{503}
\end{equation}
In gauge theory, this conservation law is exemplified by the Noether
identities.

Let
\[\tau=\tau^\la\dr_\la\]
be a vector field on $X$ and
\[ u=\tau_\G=\tau^\m (\dr_\m+\G^i_\m\dr_i)\]
its horizontal lift onto the fibred manifold $Y$ by a connection $\G$
on $Y$. In this case, the equality (\ref{502}) takes the form
\begin{equation}
\ol s^*{\bf L}_{\ol\tau_\G}L=
-\frac{d}{dx^\la}[\tau^\m T_\G{}^\la{}_\m (\ol s)]\om \label{504}
\end{equation}
where
\begin{equation}
T_\G{}^\la{}_\m (\ol s) =\pi^\la_i(\ol s^i_\m-\G^i_\m)
-\dl^\la_\m\cL \label{84}
\end{equation}
is the canonical energy-momentum
tensor of a field $\ol s$ with respect to the connection $\G$ on $Y$.
The tensor (\ref{84}) is the particular case of the
stress energy-momentum tensors \cite{fer,got2,kij}.

In particular, when the fibration $Y\to X$ is trivial,
one can choose the trivial connection $\G^i_\mu=0$.
In this case, the temsor (\ref{84}) is precisely the standard canonical
energy-momentum tensor, and if
\[{\bf L}_\tau\cL=0\]
for all vector fields $\tau$ on $X$ (e.g., $X$ is the Minkowski space),
the conservation law (\ref{504})
comes to the well-known conservation law
\[\frac{d}{dx^\la} T^\la{}_\m (\ol s)=0\]
of the canonical energy-momentum tensor.

In general, the Lie derivative
${\bf L}_{\ol\tau_\G}L$ fails to be equal to zero as a rule, and the equality
(\ref{504}) is not the conservation law of a canonical
energy-momentum tensor. For instance, in gauge theory of gauge
potentials and scalar matter fields in the presence of a background
world metric $g$, we get the covariant conservation law
\begin{equation}
\nabla_\la t^\la{}_\m=0 \label{E2}
\end{equation}
of the metric energy-momentum tensor.

In Einstein's General Relativity, the covariant conservation law (\ref{E2})
issues directly from
gravitational equations. But it is concerned only with zero-spin matter in the
presence of the gravitational field generated by this matter itself.
The total energy-momentum conservation law for matter and gravity is
introduced by hand. It reads
\begin{equation}
\frac{d}{dx^\m}[(-g)^N(t^{\la\m} +T_g{}^{\la\m)}]=0\label{E3}
\end{equation}
where the energy-momentum pseudotensor $T_g{}^{\la\m}$ of a metric
gravitational field is defined to satisfy the relation
\[
(-g)^N(t^{\la\m} +T_g{}^{\la\m})= \frac{1}{2\kappa}\dr_\si\dr_\al
[(-g)^N(g^{\la\m} g^{\si\al}-g^{\si\m} g^{\la\al})
\]
on solutions of the Einstein equations. The conservation law (\ref{E3})
is rather satisfactory only in cases  of asymptotic-flat gravitational
fields and a background gravitational field. The
energy-momentum conservation law in the affine-metric gravitation theory
and the gauge gravitation theory was not discussed widely \cite{heh}.

Thus, the Lagrangian formulation of field theory does not provide us
with any universal procedure in order to analize the
energy-momentum conservation law in gravitation theory, otherwise the
covariant multimomentum Hamiltonian formalism.
In the framerwork of this formalism, we get the
fundamental identity (\ref{5.27})
whose restriction to the Lagrangian constraint space can be treated
the energy-momentum transformation law in field theory \cite{sard,9sar}.

Lagrangian densities of field models are almost always degenerate
and the corresponding Euler-Lagrange
equations are underdetermined. To describe constraint field systems,
the multimomentum Hamiltonian formalism can be utilized
\cite{6sar,7sar,lsar}.
In the framework of this formalism,
the finite-dimensional phase space of fields is the Legendre bundle
\begin{equation}
\Pi=\op\w^n T^*X\op\otimes_Y TX\op\otimes_Y V^*Y \label{00}
\end{equation}
over $Y$ into which the Legendre
morphism $\wh L$ associated with a Lagrangian density
$L$ on $J^1Y$ takes its values. This phase space is
provided with  the fibred coordinates $(x^\la ,y^i,p^\la_i)$ such that
\[ (x^\m,y^i,p^\m_i)\circ\wh L=(x^\m,y^i,\pi^\m_i).\]
The Legendre bundle (\ref{00}) carries the multisymplectic form
\begin{equation}
\Om =dp^\la_i\w
dy^i\w\om\otimes\dr_\la. \label{406}
\end{equation}
In case of $X={\bf R}$, this form recovers
the standard symplectic form in analytical mechanics.

Building on the multisymplectic form $\Om$, one can develop
the so-called multimomentum Hamiltonian formalism of field theory where
canonical momenta correspond to derivatives of fields with respect to all world
coordinates, not only the temporal one. On the mathematical level, this is
the multisymplectic
generalization of the standard Hamiltonian formalism in analytical mechanics
to fibred manifolds over an $n$-dimensional base $X$, not only ${\bf R}$.
We say that a connection $\g$ on the fibred Legendre manifold $\Pi\to
X$ is a Hamiltonian connection if the form $\g\rfloor\Om$ is closed.
Then, a Hamiltonian form $H$ on $\Pi$ is defined to be an
exterior form such that
\begin{equation}
dH=\g\rfloor\Om \label{013}
\end{equation}
for some Hamiltonian connection $\g$. Every  Hamiltonian form admits splitting
\begin{equation}
H =p^\la_idy^i\w\om_\la -p^\la_i\G^i_\la\om
-\wt{\cH}_\G\om=p^\la_idy^i\w\om_\la-\cH\om,\qquad
\om_\la=\dr_\la\rfloor\om,  \label{017}
\end{equation}
where $\G$ is a connection on $Y\to X$.
Given the  Hamiltonian form $H$ (\ref{017}), the equality
(\ref{013}) comes to the Hamilton equations
\begin{equation}
\dr_\la y^i(x) =\dr^i_\la\cH, \qquad
\dr_\la p^\la_i(x)=-\dr_i\cH \label{3.11}
\end{equation}
for sections of the fibred Legendre manifold $\Pi\to X$.

The Hamilton equations (\ref{3.11}) are the multimomentum generalization
of the standard Hamilton equations in mechanics.
The energy-momentum transformation law (\ref{5.27}) which we suggest is
accordingly
the multimomentum generalization of the conventional energy transformation law
(\ref{E1}).
Its application to the Hamiltonian gauge theory
in the presence of a background world metric recovers the familiar metric
energy-momentum transformation law (\ref{E2}) \cite{sard,9sar}.

The identity (\ref{5.27}) remains true also in the gravitation theory.
The tools are now at hand to examine the energy-momentum transformation
for gravity. In this work, we restrict our consideration to the
affine-metric gravitation theory. The key point consists in
the feature of a metric gravitational field whose
canonical momenta on the constraint space are equal to zero \cite{8sar}.

\section{Technical preliminary}

A fibred manifold
\[\pi:Y\to X\]
is provided with fibred coordinates $(x^\la, y^i)$ where $x^\la$ are
coordinates of the base $X$.
A locally trivial fibred manifold is termed the bundle.
We denote by $VY$ and $V^*Y$ the vertical tangent bundle and the
vertical cotangent bundle of $Y$ respectively.
For the sake of simplicity, the pullbacks
$Y\op\times_XTX$ and $Y\op\times_XT^*X$
are denoted by $TX$ and $T^*X$ respectively.

On fibred manifolds, we
consider  the following types of differential forms:

(i) exterior horizontal forms $ Y\to\op\w^r T^*X$,

(ii) tangent-valued horizontal forms $Y\to\op\w^r T^*X\op\otimes_Y TY$
and, in particular, soldering forms $Y\to T^*X\op\otimes_YVY$,

(iii) pullback-valued forms
\[
Y\to \op\w^r T^*Y\op\otimes_Y TX, \qquad
Y\to \op\w^r T^*Y\op\otimes_Y T^*X.
\]
Horizontal $n$-forms are called horizontal densities.

Given a fibred manifold $Y\to X$, the first order jet manifold $J^1Y$ of
$Y$ is both the fibred manifold $J^1Y\to X$
and the affine bundle $J^1Y\to Y $  modelled on the vector
bundle $T^*X\otimes_Y VY.$

We identify $J^1Y$ to its image under the  canonical bundle monomorphism
\ben &&\la:J^1Y\op\to_YT^*X \op\otimes_Y TY,\nonumber\\
&&\la=dx^\la\otimes(\dr_\la+y^i_\la \dr_i).\label{18}\een
Given a fibred morphism of $\Phi: Y\to Y'$
over a diffeomorphism of $X$, its jet prolongation
$J^1\Phi:J^1Y\to J^1Y'$ reads
\[ {y'}^i_\m\circ
J^1\Phi=(\dr_\la\Phi^i+\dr_j\Phi^iy^j_\la)\frac{\dr x^\la}{\dr {x'}^\m}.\]

Every  vector field
\[u = u^\la\dr_\la + u^i\dr_i\]
on a fibred manifold $Y\to X$ gives rise to the projectable vector field
\begin{equation}
\ol u =u^\la\dr_\la + u^i\dr_i + (\dr_\la u^i+y^j_\la\dr_ju^i
- y_\m^i\dr_\la u^\m)\dr_i^\la, \label{1.21}
\end{equation}
on the fibred jet manifold $J^1Y\to X$ where $J^1TY$ is the jet
manifold of the fibred manifold $TY\to X$.

The canonical morphism (\ref{18}) gives rise to the bundle
monomorphism
\[
 \wh\la: J^1Y\op\times_X TX\ni\dr_\la\mapsto\wh{\dr}_\la = \dr_\la\rfloor
\la\in J^1Y\op\times_Y TY,
\qquad \wh{\dr}_\la=\dr_\la + y^i_\la \dr_i.
\]
This morphism determines the canonical horizontal
splitting of the pullback
\begin{equation}
J^1Y\op\times_Y TY=\wh\la(TX)\op\oplus_{J^1Y} VY,\label{1.20}
\end{equation}
\[
\dot x^\la\dr_\la
+\dot y^i\dr_i =\dot x^\la(\dr_\la +y^i_\la\dr_i) + (\dot y^i-\dot x^\la
y^i_\la)\dr_i.
\]
In other words, over $J^1Y$, we have the canonical horizontal splitting of
the tangent bundle $TY$.

Building on the canonical splitting (\ref{1.20}),
one gets the corresponding horizontal splittings of
a projectable vector field
\begin{equation}
u =u^\la\dr_\la +u^i\dr_i=u_H +u_V =u^\la (\dr_\la +y^i_\la
\dr_i)+(u^i - u^\la y^i_\la)\dr_i \label{31}
\end{equation}
on a fibred manifold $Y\to X$.

Given a fibred  manifold  $Y\to X$, there is
the 1:1 correspondence between the connections on $Y\to X$
and global sections
\[\G =dx^\la\otimes(\dr_\la+\G^i_\la\dr_i)\]
of the affine jet bundle $J^1Y\to Y$. Substitution of such a global
section $\G$ into the canonical horizontal splitting (\ref{1.20})
recovers the familiar horizontal splitting of the tangent bundle $TY$
with respect to the connection $\G$ on $Y$. These global sections form the
affine space modelled on the linear space of soldering forms on $Y$.

Every connection $\G$ on $Y\to X$ yields the first order differential
operator
\be && D_\G:J^1Y\op\to_YT^*X\op\otimes_YVY,\\
&&D_\G=(y^i_\la-\G^i_\la)dx^\la\otimes\dr_i,\ee
on $Y$ which is called the covariant differential relative to the
connection $\G$.

The repeated jet manifold
$J^1J^1Y$, by definition, is the first order jet manifold of
$J^1Y\to X$. It is provided with the adapted coordinates
$(x^\la ,y^i,y^i_\la ,y_{(\m)}^i,y^i_{\la\m})$.
Its subbundle $ \wh J^2Y$ with $y^i_{(\la)}= y^i_\la$ is called the
sesquiholonomic jet manifold.
The second order jet manifold $J^2Y$ of $Y$ is the subbundle
of $\wh J^2Y$ with $ y^i_{\la\m}= y^i_{\m\la}.$

\section{Lagrangian formalism}

Let $Y\to X$ be a fibred manifold and $ L=\cL\om$ a Lagrangian density
on $J^1Y$.  With $L$, the jet manifold $J^1Y$  carries
the uniqie associated Poincar\'e-Cartan form
\begin{equation}
\Xi_L=\pi^\la_idy^i\w\om_\la -\pi^\la_iy^i_\la\om +\cL\om \label{303}
\end{equation}
and the Lagrangian  multisymplectic form
\[\Om_L=(\dr_j\pi^\la_idy^j+\dr^\m_j\pi^\la_idy^j_\m)\w
dy^i\w\om\otimes\dr_\la.\]
Using the pullback of these forms onto the
repeated jet manifold $J^1J^1Y$, one can construct the exterior form
\begin{equation}
\La_L=d\Xi_L-\la\rfloor\Om_L=[y^i_{(\la)}-y^i_\la)d\pi^\la_i +
(\dr_i-\wh\dr_\la\dr^\la_i)\cL dy^i]\w\om,\label{304}
\end{equation}
\[ \la=dx^\la\otimes\wh\dr_\la,\qquad
\wh\dr_\la =\dr_\la +y^i_{(\la)}\dr_i+y^i_{\m\la}\dr^\m_i,\]
on $J^1J^1Y$.
Its restriction to the second order jet manifold $J^2Y$ of $Y$ reproduces
the familiar variational Euler-Lagrange operator
\begin{equation}
\cE_L= [\dr_i-
(\dr_\la +y^i_\la\dr_i+y^i_{\m\la}\dr^\m_i)\dr^\la_i]\cL dy^i\w\om.\label{305}
\end{equation}
The restriction of the form (\ref{304}) to the sesquiholonomic jet manifold
$\wh J^2Y$ defines the sesquiholonomic extension
$\cE'_L$  of the Euler-Lagrange operator (\ref{305}).
It is given by the expression
(\ref{305}), but with nonsymmetric coordinates $y^i_{\m\la}$.

Let $\ol s$ be a section of the fibred jet manifold $J^1Y\to X$ such that
its first order jet prolongation  $J^1\ol s$ takes its values into
$\Ker\cE'_L$. Then, $\ol s$ satisfies the first order
differential Euler-Lagrange equations (\ref{306}).
They are equivalent to the second order Euler-Lagrange equations
\begin{equation}
\dr_i\cL-(\dr_\la+\dr_\la s^j\dr_j
+\dr_\la\dr_\m s^j \dr^\m_j)\dr^\la_i\cL=0.\label{2.29}
\end{equation}
for sections $s$ of $Y$ where $\ol s=J^1s$.

We have the following conservation laws on solutions of the first order
Euler-Lagrange equations.

Let
\[ u=u^\m\dr_\m + u^i\dr_i\]
be a vector field on a fibred manifold $Y$ and $\ol u$ its jet lift
(\ref{1.21}) onto the fibred jet manifold $J^1Y\to X$. Given a
Lagrangian density $L$ on $J^1Y$, let us compute the Lie derivative
${\bf L}_{\ol u}L$. We have
\begin{equation}
{\bf L}_{\ol u}L= [\wh \dr_\la(\pi^\la_i(u^i-u^\m y^i_\m ) +u^\la\cL
)+ (u^i-u^\m y^i_\m )(\dr_i-\wh\dr_\la\dr^\la_i)\cL]\om, \label{501}
\end{equation}
\[\wh\dr_\la =\dr_\la +y^i_\la\dr_i+y^i_{\m\la}\dr^\m_i.\]
On solutions $\ol s$ of the first order Euler-Lagrange equations, the
equality (\ref{501}) comes to the conservation law (\ref{502}).

In particular, let
\[ u=\tau_\G=\tau^\m (\dr_\m+\G^i_\m\dr_i)\]
be the horizontal lift of a vector field
\[\tau=\tau^\la\dr_\la\]
on $X$ onto the fibred manifold $Y$ by a connection $\G$
on $Y$. In this case, the equality (\ref{502}) takes the form (\ref{504})
where $T_\G{}^\la{}_\m (\ol s)$ (\ref{84}) are coefficients of the
$T^*X$-valued form
\begin{equation}
T_\G(\ol s)=-(\G\rfloor\Xi_L)\circ\ol s =[\pi^\la_i(\ol s^i_\m-\G^i_\m)
-\dl^\la_\m\cL]dx^\m\otimes\om_\la \label{S14}
\end{equation}
on $X$. One can think on this form as being the
canonical energy-momentum
tensor of a field $\ol s$ with respect to the connection $\G$ on $Y$.

\section{Multimomentum Hamiltonian formalism}

Let $\Pi$ be the Legendre bundle (\ref{00}) over a fibred manifold
$Y\to X$. It is provided with the fibred coordinates $( x^\la ,y^i,p^\la_i)$:
\[{p'}^\la_i = J \frac{\dr y^j}{\dr{y'}^i} \frac{\dr
{x'}^\la}{\dr x^\m}p^\m_j, \qquad J^{-1}=\det (\frac{\dr {x'}^\la}{\dr
x^\m}). \]
By $J^1\Pi$ is meant the first order jet manifold of
$\Pi\to X$. It is coordinatized by
\[( x^\la ,y^i,p^\la_i,y^i_{(\m)},p^\la_{i\m}).\]

We call by a momentum morphism any bundle morphism
$\Phi:\Pi\to J^1Y$ over $Y$.
For instance, let $\G$ be a connection on $Y$. Then, the
composition $\wh\G=\G\circ\pi_{\Pi Y}$ is a momentum morphism. Conversely,
every momentum morphism $\Phi$ determines
the associated connection $ \G_\Phi =\Phi\circ\wh 0_\Pi$
on $Y\to X$ where $\wh 0_\Pi$ is the global zero section of $\Pi\to Y$.
Every connection $\G$ on $Y$ gives rise to the connection
\begin{equation}
\wt\G =dx^\la\otimes[\dr_\la +\G^i_\la (y)\dr_i +
(-\dr_j\G^i_\la (y)  p^\m_i-K^\m{}_{\n\la}(x) p^\n_j+K^\al{}_{\al\la}(x)
p^\m_j)\dr^j_\m]  \label{404}
\end{equation}
on $\Pi\to X$ where $K$ is a linear symmetric connection  on $T^*X$.

The Legendre manifold $\Pi$ carries the multimomentum Liouville form
\[\th =-p^\la_idy^i\w\om\otimes\dr_\la \]
and the multisymplectic form $\Om$ (\ref{406}).

The Hamiltonian formalism in fibred manifolds is formulated
intrinsically in terms of Hamiltonian connections which play the
role similar to that of Hamiltonian vector fields in the symplectic geometry.

We say that a  connection
$\g$ on the fibred Legendre manifold $\Pi\to X$ is a Hamiltonian
connection if the exterior form $\g\rfloor\Om$  is closed.
An exterior $n$-form $H$ on the
Legendre manifold $\Pi$ is called a  Hamiltonian form if
there exists a Hamiltonian connection  satisfying the equation (\ref{013}).

Let $H$ be a Hamiltonian form. For any exterior horizontal density
$\wt H=\wt{\cH}\om$ on $\Pi\to X$, the form $H-\wt H$ is a Hamiltonian form.
Conversely, if $H$ and $H'$ are  Hamiltonian forms,
their difference $H-H'$ is an exterior horizontal density on $\Pi\to X$.
Thus, Hamiltonian  forms constitute an affine space
modelled on a linear space of the exterior horizontal densities on
$\Pi\to X$.

Let $\G$ be a connection on $Y\to X$ and $\wt\G$ its lift
(\ref{404}) onto $\Pi\to X$. We have the equality
\[\wt\G\rfloor\Om =d(\wh\G\rfloor\th).\]
A glance at this equality shows that $\wt\G$ is a Hamiltonian
connection and
\[ H_\G=\wh\G\rfloor\th =p^\la_i dy^i\w\om_\la -p^\la_i\G^i_\la\om\]
is a Hamiltonian form. It follows that every
Hamiltonian form on $\Pi$ can be
given by the expression (\ref{017}) where $\G$ is some
connection on $Y\to X$.
Moreover, a Hamiltonian form has the canonical splitting (\ref{017})
as follows.
Given a  Hamiltonian form $H$, the vertical tangent morphism
$VH$ yields the momentum morphism
\[ \wh H:\Pi\to J^1Y, \qquad y_\la^i\circ\wh H=\dr^i_\la\cH,\]
and the associated connection $\G_H =\wh H\circ\wh 0$
on $Y$. As a consequence, we have the canonical splitting
\[ H=H_{\G_H}-\wt H.\]

The Hamilton operator $\cE_H$ for a Hamiltonian form $H$
is defined to be the first order differential operator
\begin{equation}
\cE_H=dH-\wh\Om=[(y^i_{(\la)}-\dr^i_\la\cH) dp^\la_i
-(p^\la_{i\la}+\dr_i\cH) dy^i]\w\om, \label{3.9}
\end{equation}
where $\wh\Om$
is the pullback of the multisymplectic form $\Om$ onto $J^1\Pi$.

For any connection $\g$ on $\Pi\to X$, we have
\[\cE_H\circ\g =dH-\g\rfloor\Om.\]
It follows that  $\g$  is a Hamiltonian jet field for a
Hamiltonian form $H$ if and only if it takes its values into
$\Ker\cE_H$, that is, satisfies  the algebraic Hamilton equations
\begin{equation}
\g^i_\la =\dr^i_\la\cH, \qquad \g^\la_{i\la}=-\dr_i\cH. \label{3.10}
\end{equation}

Let a Hamiltonian connection $\g$ has an integral section $r$ of
$\Pi\to X$, that is, $\g\circ r=J^1r$.
Then, the Hamilton equations (\ref{3.10}) are brought into the first
order differential Hamilton equations (\ref{3.11}).

Now we consider relations between Lagrangian and Hamiltonian
formalisms on fibred manifolds in case of semiregular Lagrangian densities
$L$ when the preimage $\wh L^{-1}(q)$ of each point of
$q\in Q$ is the connected submanifold of $J^1Y$.

Given a Lagrangian density $L$, the vertical tangent morphism $VL$ of
$L$ yields the Legendre morphism
\[
\wh L : J^1Y\to \Pi, \qquad
p^\la_i\circ\wh L =\pi^\la_i.\]

We say that a  Hamiltonian form
$H$ is associated with a Lagrangian density $L$ if $H$ satisfies the relations
\bea &&\wh L\circ\wh H\mid_Q=\Id_Q, \qquad Q=\wh L( J^1Y) \label{2.30a},\\
&& H=H_{\wh H}+L\circ\wh H, \label{2.30b}\eea
or in the coordinate form
\[
\dr^\mu_i\cL(x^\la, y^j, \dr^j_\la\cH)= p^\mu_i, \qquad p^\m_i\in Q,
\]
\[
\cL(x^\la, y^j, \dr^j_\la\cH)=p^\m_i\dr^i_\m\cH -\cH.
\]
Notete that different  Hamiltonian forms can be associated with the same
Lagrangian density.

Let a  section $r$ of $\Pi\to X$
be a solution of the Hamilton equations (\ref{3.11})
for a Hamiltonian form $H$ associated with a semiregular Lagrangian
density $L$. If $r$ lives on the constraint space $Q$, the section
$\ol s=\wh H\circ r$ of $J^1Y\to X$ satisfies the first
order Euler-Lagrange equations (\ref{306}).
Conversely, given a semiregular Lagrangian density $L$, let
$\ol s$ be a solution of the
first order Euler-Lagrange equations (\ref{306}).
Let $H$ be a Hamiltonian form associated with $L$ so that
\begin{equation}
\wh H\circ \wh L\circ \ol s=\ol s.\label{2.36}
\end{equation}
Then, the section $r=\wh L\circ \ol s$ of $\Pi\to X$ is a solution of the
Hamilton equations (\ref{3.11}) for $H$.
For sections $\ol s$ and $r$, we have the relations
\[\ol s=J^1s, \qquad  s=\pi_{\Pi Y}\circ r\]
where $s$ is a solution of the second order Euler-Lagrange equations
(\ref{2.29}).

We shall say that a family of Hamiltonian forms $H$
associated with a semiregular Lagrangian density $L$ is
complete if, for each solution $\ol s$ of the first order Euler-Lagrange
equations (\ref{306}), there exists
a solution $r$ of the Hamilton equations (\ref{3.11}) for
some  Hamiltonian form $H$ from this family so that
\begin{equation}
r=\wh L\circ\ol s,\qquad  \ol s =\wh H\circ r, \qquad
\ol s= J^1(\pi_{\Pi Y}\circ r). \label{2.37}
\end{equation}
Such a complete family
exists iff, for each solution $\ol s$ of the Euler-Lagrange
equations for $L$, there exists a  Hamiltonian form $H$ from this
family so that the condition (\ref{2.36}) holds.

The most of field models possesses affine and
quadratic Lagrangian densities. Complete
families of Hamiltonian forms associated with such Lagrangian densities
always exist \cite{sard,7sar}.

\section{Energy-momentum conservation laws}

In the framerwork of the multimomentum Hamiltonian formalism, we get the
fundamental identity whose restriction to the Lagrangian constraint
space recovers the familiar energy-momentum conservation law \cite{sard,9sar}.

Let $H$ be a Hamiltonian form on the Legendre bundle $\Pi$
(\ref{00}) over a fibred manifold $Y\to X$. Let $r$ be a section of
of the fibred Legendre manifold $\Pi\to X$ and $(y^i(x), p^\la_i(x))$ its
local components. Given a connection $\G$ on $Y\to X$, we consider the
following $T^*X$-valued $(n-1)$-form on $X$:
\ben &&T_\G(r)=-(\G\rfloor H)\circ r,\nonumber\\
&&T_\G (r)=[p^\la_i(y^i_\m -\G^i_\m)-\dl^\la_\m(p^\al_i(y^i_\al-\G^i_\al)
-\wt{\cH}_\G)] dx^\m\otimes\om_\la, \label{5.8}\een
where $\wt{\cH}_\G$ is the Hamiltonian density in the splitting
(\ref{017}) of $H$ with respect to the connection $\G$.

Let
\[\tau=\tau^\la\dr_\la\]
be a vector field on $X$. Given a connection $\G$ on $Y\to X$, it
gives rise to the vector field
\[\wt\tau_\G= \tau^\la\dr_\la + \tau^\la\G^i_\la\dr_i +
(-\tau^\m p^\la_j\dr_i\G^j_\m
-p^\la_i\dr_\m\tau^\m + p^\m_i\dr_\m\tau^\la) \dr^i_\la\]
on the Legendre bundle $\Pi$. Let us calculate the Lie derivative
${\bf L}_{\wt\tau_\G}\wt H_\G$ on a section $r$. We have
\begin{equation}
({\bf L}_{\wt\tau_\G}\wt H_\G)\circ r=p^\la_iR^i_{\la\m}+d[\tau^\m
T_\G{}^\la{}_\m (r)\om_\la]-(\wt\tau_{\G V}\rfloor\cE_H)\circ r\label{221}
\end{equation}
where
\be &&R =\frac12 R^i_{\la\m} dx^\la\wedge dx^\m\otimes\dr_i=\\
&&\quad \frac12 (\dr_\la\G^i_\m -\dr_\m\G^i_\la +\G^j_\la\dr_j\G^i_\m
-\G^j_\m\dr_j\G^i_\la) dx^\la\wedge dx^\m\otimes\dr_i; \ee
is the curvature of  the connection $\G$,  $\cE_H$ is the Hamilton
operator (\ref{3.9}) and
\[\wt\tau_{\G V}=\tau^\la(\G^i_\la-y^i_\la)\dr_i +
(-\tau^\m p^\la_j\dr_i\G^j_\m
-p^\la_i\dr_\m\tau^\m+p^\m_i\dr_\m\tau^\la-\tau^\m p^{\la\m}_i) \dr^i_\la\]
is the vertical part of the canonical
horizontal splitting (\ref{31}) of the vector field $\wt\tau_\G$ on $\Pi$
over $J^1\Pi$. If $r$ is a
solution of the Hamilton equations, the equality (\ref{221}) comes
to the identity
\begin{equation}
(\dr_\m+\G^i_\m\dr_i-\dr_i\G^j_\m
p^\la_j\dr^i_ \la)\wt{\cH}_\G=\frac{d}{dx^\la}
T_\G{}^\la{}_\m (r)+ p^\la_iR^i_{\la\m}. \label{5.27}
\end{equation}
On solutions of the Hamilton equations, the form (\ref{5.8}) reads
\begin{equation}
T_\G(r)=[p^\la_i\dr^i_\m\wt{\cH}_\G-
\dl^\la_\m(p^\al_i\dr^i_\al\wt{\cH}_\G-\wt{\cH}_\G)]
dx^\m\otimes\om_\la.\label{5.26}
\end{equation}
One can verify that the identity (\ref{5.27}) does not depend upon choice
of the connection $\G$.

For instance,
if $X={\bf R}$ and $\G$ is the trivial connection, then
\[
T_\G(r)=\cH dt
\]
where $\cH$ is a Hamiltonian function and the identity
(\ref{5.27}) consists with the familiar energy transformation law (\ref{E1}).

To clarify the physical
meaning of (\ref{5.27}) when $n>1$,
we turn to the Lagrangian formalism.
Let a multimomentum Hamiltonian form $H$ be associated with a
semiregular Lagrangian density $L$. Let $r$ be a solution
of the Hamilton equations for $H$ which lives on the Lagrangian
constraint space $Q$ and $\ol s$ the associated solution  of the
first order
Euler-Lagrange equations for $L$ so that $r$ and $\ol s$ satisfy the conditions
(\ref{2.37}). Then, we have
\[
T_\G (r)= T_\G(\ol s)
\]
where $T_\G(\ol s)$
is the Lagrangian canonical energy-momentum tensor (\ref{S14}).
It follows that the form (\ref{5.26}) may be treated as a Hamiltonian
canonical energy-momentum tensor with respect to a background connection
$\G$ on the fibred manifold $Y\to X$ (or a Hamiltonian
stress energy-momentum tensor).
At the same time, the identity (\ref{5.27}) in gauge theory turns out
to be precisely the covariant
energy-momentum conservation law
for the metric energy-momentum tensor, not the canonical one \cite{9sar}.

In the Lagrangian formalism, the metric
energy-momentum tensor is defined to be
\[ \sqrt{-g} t_{\al\bt}=2\frac{\dr\cL}{\dr g^{\al\bt}}.\]
In case of a background world metric $g$, this object is well-behaved.
In the framework of the multimomentum Hamiltonian formalism,
one can introduce the similar tensor
\begin{equation}
\sqrt{-g}t_H{}^{\al\bt}=2\frac{\dr\cH}{\dr g_{\al\bt}}.\label{5.28}
\end{equation}

Recall the useful relation
\[
\frac{\dr}{\dr g^{\al\bt}} = -g_{\al\m}g_{\bt\n}\frac{\dr}{g_{\m\n}}.
\]

If a multimomentum Hamiltonian form $H$ is associated
with a semiregular Lagrangian density $L$, we have the equalities
\be
&&
t_H{}^{\al\bt}(q)
=-g^{\al\m}g^{\bt\n}t_{\m\n}(x^\la,y^i,\dr_\la^i\cH(q)), \\
&&  t_H{}^{\al\bt}(x^\la,y^i,\pi^\la_i(z))=-g^{\al\m}g^{\bt\n}t_{\m\n}(z)
\ee
where $q\in Q$, $z\in J^1Y$ and
\[ \wh H\circ\wh L(z)=z.\]
In view of these equalities, we can think of the tensor (\ref{5.28})
restricted to the Lagrangian constraint space $Q$ as being the
Hamiltonian metric energy-momentum tensor.
On $Q$, the tensor (\ref{5.28}) does not depend upon choice of
a Hamiltonian form $H$ associated with $L$. Therefore, we shall
denote it by the common symbol $t$. Set
\[
t^\la{}_\al = g_{\al\n}t^{\la\n}.
\]

In the presence of a background world metric
$g$, the identity (\ref{5.27}) takes the form
\begin{equation}
t^\la{}_\al\{^\al{}_{\la\m}\}\sqrt{-g}
+(\G^i_\m\dr_i-\dr_i\G^j_\m p^\la_j\dr^i_
\la)\wt{\cH}_\G=\frac{d}{dx^\la} T_\G{}^\la{}_\m +p^\la_iR^i_{\la\m}
\label{5.29}
\end{equation}
where
\[\frac{d}{dx^\la} = \dr_\la +\dr_\la y^i\dr_i +\dr_\la p^\m_i\dr_\m^i\]
and by $\{^\al{}_{\la\m}\}$ are meant the Cristoffel symbols of the
world metric $g$.

When applied to the Hamiltonian gauge theory
in the presence of a background world metric, the identity (\ref{5.29})
recovers the familiar metric
energy-momentum transformation law (\ref{E2}) \cite{sard,9sar}.

\section{Energy-momentum conservation laws \newline
in affine-metric gravitation theory}

The contemporary concept of  gravitation interaction is based on
the gauge gravitation theory with  two types of gravitational
fields. These are tetrad garvitational fields and Lorentz gauge potentials.
At present, all Lagrangian densities of
 classical and quantum gravitation
theories are expressed in these variables. They are of the first order
with respect to these fields.
Only General Relativity without fermion
matter sources utilizes traditionally  the Hilbert-Einstein Lagrangian
density $L_{HE}$ which is of the second order  with respect
to a pseudo-Riemannian metric. One can reduce its order by means of
the Palatini variables when the Levi-Civita
connection is regarded on the same footing as a pseudo-Riemannian
metric.

Here we consider the
affine-metric gravitation theory when there is no fermion matter and
gravitational variables are both a pseudo-Riemannian metric $g$ on a world
manifold $X^4$ and linear connections $K$
on the tangent bundle of $X^4$. We
call them a world metric and a world connection respectively.
Given a world metric, every
world connection meets the well-known decomposition in the Cristoffel
symbols, contorsion and the nonmetricity term.
We here are not concerned with
the matter interecting with a general linear connection, for it
is non-Lagrangian and hypothetical as a rule.

In the rest of the article, $X$ is an oriented 4-dimensional
world manifold which obeys the
well-known topological conditions in order that a gravitational field
exists on $X^4$.

Let $LX\to X^4$ be the principal bundle of linear frames in the
tangent spaces to $X^4$. The structure group of $LX$ is the group
\[
GL_4=GL^+(4,{\bf R})
\]
of general linear transfromations of the standard fibre ${\bf R}^4$ of
the tangent bundle $TX$.
The world connections are associated with the principal connections on the
principal bundle $LX\to X^4$.
 Hence, there is the 1:1 correspondence between the
world connections and the global sections of the principal connection
bundle
\begin{equation}
C=J^1LX/GL_4. \label{251}
\end{equation}
 Therefore, we can apply the standard procedure of the Hamiltonian
gauge theory in order to describe the configuration and phase spaces of
world connections \cite{7sar,lsar,8sar}.

 There is the 1:1 correspondence between the world metrics $g$ on
$X^4$  and the global sections of the bundle $\Sigma_g$ of
pseudo-Riemannian bilinear
forms in tangent spaces to $X^4$. This bundle is
associated with the $GL_4$-principal bundle $LX$.
The 2-fold covering of the bundle $\Si_g$ is the quotient bundle
\begin{equation}
\Sigma=LX/SO(3,1)\label{5.15}
\end{equation}
where by $SO(3,1)$ is meant the connected Lorentz group.

Thus, the total configuration space of the
affine-metric gravitational variables is represented by the
product of the corresponding jet manifolds:
\begin{equation}
J^1C\op\times_{X^4}J^1\Sigma. \label{N33}
\end{equation}
Given a
holonomic bundle atlas of $LX$ associated with  induced coordinates of
$TX$ and $T^*X$, this
 configuration space is provided with the adapted coordinates
\[
(x^\m, g^{\al\bt}, k^\al{}_{\bt\m}, g^{\al\bt}{}_\la,
k^\al{}_{\bt\m\la}).
\]

Also the total phase space $\Pi$ of the affine-metric gravity
is the product of the Legendre bundles
over the above-mentioned bundles $C$ and $\Si$.
It is coordinatized by the corresponding canonical coordinates
\[
(x^\m, g^{\al\bt}, k^\al{}_{\bt\m},
p_{\al\bt}{}^\la, p_\al{}^{\bt\m\la}).
\]

On the configuration space (\ref{N33}), the  Hilbert-Einstein  Lagrangian
density of General Relativity reads
\begin{equation}
L_{HE}=-\frac{1}{2\kappa}g^{\bt\la}\cF^\al{}_{\bt\al\la}
\sqrt{-g}\omega,\label{5.17}
\end{equation}
 \[
\cF^\al{}_{\bt\n\la}=k^\al{}_{\bt\la\n}-
k^\al{}_{\bt\n\la}+k^\al{}_{\ve\n}
k^\ve{}_{\bt\la}-k^\al{}_{\ve\la}
k^\ve{}_{\bt\n}.
\]

The corresponding Legendre morphism is given by the expressions
\ben
&& p_{\al\bt}{}^\la\circ \wh L_{HE}=0,\nonumber
\\ &&   p_\al{}^{\bt\n\la}\circ \wh L_{HE}
=\pi_\al{}^{\bt\n\la} =\frac{1}{2\kappa}(\dl^\n_\al
g^{\bt\la}-\dl^\la_\al g^{\bt\n})\sqrt{-g}.\label{5.18}
\een
These relations define the constraint space of General Relativity  in
multimomentum canonical variables.

Building on the set of connections on the bundle $C\times\Si$,
one can construct the complete family of  multimomentum Hamiltonian
forms associated with the Lagrangian density (\ref{5.17}). To
minimize it, we consider the following subset of these connections.

Let $K$ be a world connection and
\be
&& S_K{}^\al{}_{\bt\n\la}=\frac12
[k^\al{}_{\ve\n}
k^\ve{}_{\bt\la}-k^\al{}_{\ve\la}
k^\ve{}_{\bt\n} +\dr_\la K^\al{}_{\bt\n}
+\dr_\n K^\al{}_{\bt\la}\\
&& \qquad
-2K^\ve{}_{(\n\la)}(K^\al{}_{\bt\ve}
-k^\al{}_{\bt\ve}) +
K^\ve{}_{\bt\la}k^\al{}_{\ve\n}
+K^\ve{}_{\bt\n}k^\al{}_{\ve\la}\\
&& \qquad -
K^\al{}_{\ve\la}k^\ve{}_{\bt\n}
-K^\al{}_{\ve\n}k^\ve{}_{\bt\la}]
\ee
the corresponding connection on the bundle $C$ (\ref{251}). Let $K'$ be
another symmetric world connection which induces an associated
connection on the bundle $\Si$.  On
the bundle $C\times\Si$, we consider the following connection
\ben
&& \G^{\al\bt}{}_\la =-{K'}^\al{}_{\ve\la}g^{\ve\bt} -
{K'}^\bt{}_{\ve\la} g^{\al\ve}, \nonumber\\
&&\G^\al{}_{\bt\n\la} = S_K{}^\al{}_{\bt\n\la}
-\frac12R^\al{}_{\bt\n\la} \label{N34}
 \een
where
\[
R^\al{}_{\bt\n\la}=K^\al{}_{\bt\n\la}-
K^\al{}_{\bt\la\nu}+K^\al{}_{\ve\la}
K^\ve{}_{\bt\nu}-K^\al{}_{\ve\nu}
K^\ve{}_{\bt\la}
\]
is the curvature of the connection $K$. The corresponding
multimomentum Hamiltonian form is given by the expression
 \ben
&& H_{HE}=(p_{\al\bt}{}^\la dg^{\al\bt} +
p_\al{}^{\bt\n\la}dk^\al{}_{\bt\n})\wedge\omega_\la
-\cH_{HE}\omega, \nonumber \\
&& \cH_{HE}=-p_{\al\bt}{}^\la({K'}^\al{}_{\ve\la}
g^{\ve\bt} +{K'}^\bt{}_{\ve\la}
g^{\al\ve})
+p_\al{}^{\bt\n\la}S^\al{}_{\bt\n\la}\nonumber \\
 && \qquad -\frac12R^\al{}_{\bt\n\la}
(p_\al{}^{\bt\n\la}-\pi_\al{}^{\bt\n\la}). \label{5.19}
\een
It is associated with the Lagrangian density $L_{HE}$. We shall justify
that the
multimomentum Hamiltonian forms (\ref{5.19}) parameterized by all the world
connections $K$ and $K'$ constitute the complete family.

Given the multimomentum Hamiltonian form $H_{HE}$ (\ref{5.19}) plus
that $H_M$ of matter,
the corresponding covariant Hamilton equations for General Relativity read
\bea
&&\dr_\la
g^{\al\bt} +{K'}^\al{}_{\ve\la}g^{\ve\bt}
+{K'}^\bt{}_{\ve\la}g^{\al\ve}=0, \label{5.20a}\\
&&\dr_\la k^\al{}_{\bt\n}=
S^\al{}_{\bt\n\la}
-\frac12R^\al{}_{\bt\n\la}, \label{5.20b} \\
&&\dr_\la p_{\al\bt}{}^\la =p_{\ve\bt}{}^\sigma
{K'}^\ve{}_{\al\sigma} +
p_{\ve\al}{}^\sigma {K'}^\ve{}_{\bt\sigma} \nonumber\\
&& \qquad
-\frac{1}{2\kappa}(R_{\al\bt} -\frac12g_{\al\bt}R)\sqrt{-g}
-\frac{\dr\cH_M}{\dr g^{\al\bt}},\label{5.20c} \\
&& \dr_\la p_\al{}^{\bt\n\la}
= -p_\al{}^{\ve[\n\g]}
k^\bt{}_{\ve\g}
 +p_\ve{}^{\bt[\n\g]}
k^\ve{}_{\al\g} \nonumber \\
&& \qquad -
 p_\al{}^{\bt\ve\g}
K^\n{}_{(\ve\g)} -p_\al{}^{\ve(\n\g)}
K^\bt{}_{\ve\g}
 +p_\ve{}^{\bt(\n\g)}
K^\ve{}_{\al\g}. \label{5.20d}
\eea
The Hamilton equations (\ref{5.20a}) and (\ref{5.20b}) are independent
of canonical momenta and so, reduce to the gauge-type condition.
The gauge-type
condition (\ref{5.20b}) breaks into two parts
\ben
&&\cF^\al{}_{\bt\la\nu}=R^\al{}_{\bt\n\la},\label{5.21}\\
&& \dr_\n(K^\al{}_{\bt\la} -k^\al{}_{\bt\la})
+\dr_\la(K^\al{}_{\bt\n} -k^\al{}_{\bt\n}) \nonumber\\
&&\qquad  -2K^\ve{}_{(\n\la)}
(K^\al{}_{\bt\ve} -k^\al{}_{\bt\ve}) +
K^\ve{}_{\bt\la}k^\al{}_{\ve\n}
+K^\ve{}_{\bt\n}k^\al{}_{\ve\la} \nonumber\\
&& \qquad   -K^\al{}_{\ve\la}k^\ve{}_{\bt\n}
-K^\al{}_{\ve\n}k^\ve{}_{\bt\la}=0.
\label{5.22}
\een
It is readily observed that,
for a given world metric $g$ and a world connection $k$, there
always exist the world connections $K'$ and $K$ such that the
gauge-type conditions (\ref{5.20a}), (\ref{5.21}) and (\ref{5.22}) hold
(e.g. $K'$ is the Levi-Civita connection of $g$ and $K=k$).
It follows that the multimomentum Hamiltonian forms (\ref{5.19}) consitute
really the complete family.

 Being restricted to the constraint space (\ref{5.18}), the Hamilton
equations (\ref{5.20c}) and (\ref{5.20d}) comes to
\ben
&& \frac{1}{\kappa}(R_{\al\bt}
-\frac12 g_{\al\bt}R)\sqrt{-g}= -\frac{\dr\cH_M}{\dr g^{\al\bt}},
\label{5.23} \\
&& D_\al(\sqrt{-g}g^{\n\bt}) - \dl^\n_\al
D_\la(\sqrt{-g}g^{\la\bt}) +\sqrt{-g}[g^{\n\bt}(k^\la{}_{\al\la} -
k^\la{}_{\la\al}) \nonumber\\
&& \qquad + g^{\la\bt}(k^\n{}_{\la\al}-k^\n{}_{\al\la})+ \dl^\n_\al
g^{\la\bt} (k^\m{}_{\m\la} - k^\m{}_{\la\m})] =0 \label{5.24}
\een
where
\[
D_\la g^{\al\bt}= \dr_\la g^{\al\bt} + k^\al{}_{\m\la}g^{\m\bt} +
k^\bt{}_{\m\la}g^{\al\m}.
\]
Substituting Eq.(\ref{5.21}) into Eq.(\ref{5.23}),
we obtain the Einstein equations
\begin{equation}
\frac{1}{\kappa}\cF_{\al\bt}-\frac12 g_{\al\bt}\cF= -t_{\al\bt}\label{5.25}
\end{equation}
where $t_{\al\bt}$ is the metric energy-momentum tensor of matter.
It is easy to see that Eqs.(\ref{5.24}) and (\ref{5.25}) are the
familiar equations of gravitation theory phrased in terms of
the generalized Palatini variables. In particular,
the former is the equation for
the torsion and the nonmetricity term of the connection $k^\al{}_{\bt\n}$.
In the absence of matter sources of a general linear connection,
it admits the well-known solution
\be
&&k^\al{}_{\bt\n} =\{^\al{}_{\bt\n}\} - \frac12\dl^\al_\n V_\bt,\\
&& D_\al g^{\bt\g}= V_\al g^{\bt\g}
\ee
where $V_\al$ is an arbitrary covector field corresponding to the
well-known projective freedom.

Turn now to the identity (\ref{5.27}). It takes the form
\ben
&&(\dr_\m+\G^{\al\bt}{}_\m\dr_{\al\bt}+\G^i_\m\dr_i-\dr_i\G^j_\m
p^\la_j\dr^i_ \la)( \wt{\cH}_{HE}+\wt{\cH}_M) = \nonumber\\
&& \qquad \frac{d}{dx^\la}(
T^\la{}_\m+T_M{}^\la{}_\m)+ p_\al{}^{\bt\nu\la}R^\al{}_{\bt\nu\la\mu}
 + p^\la_iR^i_{\la\m} \label{E6}
\een
where $T$ and $T_M$ are the canonical energy-momentum tensors of
affine-metric gravity and matter respectively.

The energy-momentum tensor (\ref{5.26}) of affine-metric gravity on solutions
of the Hamilton equations reads
\begin{equation}
T^\la{}_\mu = \dl^\la_\mu \wt{\cH}_{HE}=
\frac12\dl^\la_\mu\pi_\al{}^{\bt\nu\ve}R^\al{}_{\bt\nu\ve}=
\frac{1}{2\kappa}\dl^\la_\mu P\sqrt{-g}. \label{E5}
\end{equation}
It follows that
\[
(\dr_\m +\G^{\al\bt}{}_\m\dr_{\al\bt})\wt\cH_{HE}=
\frac{d}{dx^\la}T^\la{}_\mu.
\]
One can verify also that, on solutions of the Hamilton equations, the curvature
of the connection (\ref{N34}) vanishes. Then, the identity (\ref{E6})
takes the form (\ref{5.29}). In gauge theory, it
is reduced to the familiar conservation law (\ref{E2}). This is however
a very particular case because the Hilbert-Einstein Hamiltonian density
$\wt\cH_{HE}$ is independent of gravitational momenta.

Since the canonical momenta $p_{\al\bt}{}^\la$ of the
world metric are equal to zero and the Hamilton equation (\ref{5.20c})
comes to
\[
\dr_{\al\bt}( \wt{\cH}_{HE}+\wt{\cH}_M)=0,
\]
the equality (\ref{E6}) can be rewritten
\begin{equation}
\pi_\al{}^{\bt\nu\la}\dr_\m R^\al{}_{\bt\nu\la} +
(\G^i_\m\dr_i-\dr_i\G^j_\m p^\la_j\dr^i_ \la)\wt{\cH}_M=
\frac{d}{dx^\la}(T^\la{}_\m+T_M{}^\la{}_\m)+ p^\la_iR^i_{\la\m}.
\label{E7}
\end{equation}
In gauge theory, it takes the form
\begin{equation}
\pi_\al{}^{\bt\nu\la}\dr_\m R^\al{}_{\bt\nu\la} - p^\la_iR^i_{\la\m}=
\frac{d}{dx^\la}(T^\la{}_\m+T_M{}^\la{}_\m).
\label{E8}
\end{equation}
It is the form of the energy-momentum transformation law which
we observe also in the case of the quadratic Lagrangian densities of
affine-metric gravity.

As a test case, let us consider the sum
\begin{equation}
L = (-\frac{1}{2\kappa}g^{\bt\la}\cF^\al{}_{\bt\al\la} +\frac{1}{4\ve}
g_{\al\g}g^{\bt\si}g^{\nu\mu}g^{\la\ve}\cF^\al{}_{\bt\nu\la}
\cF^\g{}_{\si\mu\ve})\sqrt{-g}\om \label{E18}
\end{equation}
of the Hilbert-Einstein Lagrangian density and the Yang-Mills one.
The corresponding Legendre morphism reads
\bea
&& p_{\al\bt}{}^\la\circ\wh L=0,\label{E9a}\\
&& p_\al{}^{\bt(\nu\la)}\circ\wh L=0, \label{E9b}\\
&& p_\al{}^{\bt[\nu\la]}\circ\wh L=\pi_\al{}^{\bt\nu\la} +\frac1\ve
g_{\al\g}g^{\bt\si}g^{\nu\mu}g^{\la\ve}\cF^\g{}_{\si\ve\mu}\sqrt{-g}.
\label{E9c}
\eea

To construct the complete family of multimomentum Hamiltonian forms
associated with the Lagrangian density (\ref{E18}),
we consider the following connection
\ben
&& \G^{\al\bt}{}_\la =-{K'}^\al{}_{\ve\la}g^{\ve\bt} -
{K'}^\bt{}_{\ve\la} g^{\al\ve}, \nonumber\\
&&\G^\al{}_{\bt\n\la} = S_K{}^\al{}_{\bt\n\la} \label{E10}
\een
on the bundle $C\times\Si$ where we utilize the notations of the expression
(\ref{N34}). Then the multimomentum Hamiltonian forms
\ben
&& H=(p_{\al\bt}{}^\la dg^{\al\bt} +
p_\al{}^{\bt\n\la}dk^\al{}_{\bt\n})\w\om_\la-\cH\om, \nonumber \\
&& \cH=-p_{\al\bt}{}^\la({K'}^\al{}_{\ve\la}g^{\ve\bt} +{K'}^\bt{}_{\ve\la}
g^{\al\ve})+p_\al{}^{\bt\n\la}S^\al{}_{\bt\n\la}\nonumber \\
&& \qquad +\frac{\ve}4 g^{\al\g}g_{\bt\si}g_{\nu\mu}g_{\la\ve}
(p_\al{}^{\bt[\nu\la]}-\pi_\al{}^{\bt\nu\la}) (p_\g{}^{\si[\m\ve]}
-\pi_\g{}^{\bt\m\ve})  \label{E11}
\een
are associated with the Lagrangian density {\ref{E18}) and constitute
the complete family.

The corresponding Hamilton equations read
\bea
&&\dr_\la
g^{\al\bt} +{K'}^\al{}_{\ve\la}g^{\ve\bt}
+{K'}^\bt{}_{\ve\la}g^{\al\ve}=0, \label{E12a}\\
&&\dr_\la k^\al{}_{\bt\n}=S^\al{}_{\bt\n\la} +
\ve g^{\al\g}g_{\bt\si}g_{\nu\mu}g_{\la\ve}(p_\g{}^{\si[\m\ve]}
-\pi_\g{}^{\bt\m\ve}) , \label{E12b} \\
&&\dr_\la p_{\al\bt}{}^\la =-\frac{\dr\cH}{\dr g^{\al\bt}}
-\frac{\dr\cH_M}{\dr g^{\al\bt}},\label{E12c} \\
&& \dr_\la p_\al{}^{\bt\n\la}= -p_\al{}^{\ve[\n\g]}
k^\bt{}_{\ve\g} +p_\ve{}^{\bt[\n\g]}k^\ve{}_{\al\g} \nonumber \\
&& \qquad - p_\al{}^{\bt\ve\g}K^\n{}_{(\ve\g)} -p_\al{}^{\ve(\n\g)}
K^\bt{}_{\ve\g} +p_\ve{}^{\bt(\n\g)}K^\ve{}_{\al\g}. \label{E12d}
\eea
The equation (\ref{E12b}) breaks into the equation (\ref{E9c}) and the
equation (\ref{5.22}).
It is readily observed that,
for a given world metric $g$ and a world connection $k$, there
always exist the world connections $K'$ and $K$ such that the
gauge-type conditions (\ref{E12a}) and (\ref{5.22}) hold
(e.g. $K'$ is the Levi-Civita connection of $g$ and $K=k$).
It follows that the multimomentum Hamiltonian forms (\ref{E11}) consitute
really the complete family.

Substituting Eq.(\ref{E12b}) into Eq.(\ref{E12c}) on the constraint space
(\ref{E9a}), we get the Einstein equations. On the constraint space
(\ref{E9b}) - (\ref{E9c}), the equation (\ref{E12d}) is the Yang-Mills
generalization
\[
D_\la p_\al{}^{\bt\n\la}=\dr_\la p_\al{}^{\bt\n\la} +p_\al{}^{\ve[\n\g]}
k^\bt{}_{\ve\g} -p_\ve{}^{\bt[\n\g]}k^\ve{}_{\al\g} =0
\]
of the equation (\ref{5.24}).

Turn now to the energy-momentum conservation law. For the sake simplicity,
let us consider the splitting of the multimomentum Hamiltonian form
(\ref{E8}) with regard to the connection (\ref{N34}) where the
Hamiltonian density is
\[
\wt{\cH}_\G= \wt{\cH} +\frac12 p_\al{}^{\bt\nu\la}R^\al{}_{\bt\nu\la}.
\]
We have the relation
\[
\frac{\dr\wt\cH_\G}{\dr_\al{}^{\bt\nu\la}}= R^\al{}_{\bt\nu\la}
\]
on solutions of the Hamilton equations. The canonical energy-momentum
tensor associated with the Hamiltonian density $\wt\cH_\G$ is written
\be
&& T_\G{}^\la{}_\mu = \frac12 p_\al{}^{\bt[\nu\la]}R^\al{}_{\bt\n\m} +
\frac{\ve}{2}g^{\al\g}g_{\bt\si}g_{\nu\dl}g_{\mu\ve}
p_\al{}^{\bt[\n\la]} (p_\g{}^{\si[\dl\ve]}-\pi_\g{}^{\si\dl\ve})\\
&&\qquad -\dl^\la_\mu (\wt{\cH} +
\frac{\ve}{2}g^{\al\g}g_{\bt\si}g_{\nu\dl}g_{\tau\ve}
\pi_\al{}^{\bt\n\tau} (p_\g{}^{\si[\dl\ve]}-\pi_\g{}^{\si\dl\ve})) =\\
&&\qquad  \frac1\ve R_\al{}^{\bt\nu\la}R^\al{}_{\bt\n\m}
 + \pi_\al{}^{\bt\nu\la}R^\al{}_{\bt\n\m}
 -\dl^\la_\mu (\frac{1}{4\ve} R_\al{}^{\bt\nu\la}R^\al{}_{\bt\n\la}+
\frac{1}{2\kappa}R).
\ee
The identity (\ref{5.27}) takes the form
\be
&&(\dr_\m+\G^{\al\bt}{}_\m\dr_{\al\bt}+\G^i_\m\dr_i-\dr_i\G^j_\m
p^\la_j\dr^i_ \la-p_\al{}^{\bt\nu\la} \frac{\dr}{\dr k^\si{}_{\g\dl}}
S^\al{}_{\bt\nu\mu}\frac{\dr}{\dr P_\si{}^{\g\dl\la}})
(\wt{\cH}_\G+\wt{\cH}_M) = \\
&& \qquad \frac{d}{dx^\la}(
T_\G{}^\la{}_\m+T_M{}^\la{}_\m)+ p_\al{}^{\bt\nu\la}R^\al{}_{\bt\nu\la\mu}
 + p^\la_iR^i_{\la\m}
\ee
On solutions of the Hamilton equations, we have
\[
(\G^i_\m\dr_i-\dr_i\G^j_\m p^\la_j\dr^i_ \la)\wt{\cH}_M
-p_\al{}^{\bt\nu\la} \frac{\dr}{\dr k^\si{}_{\g\dl}}
S^\al{}_{\bt\nu\mu}\frac{\dr}{\dr p_\si{}^{\g\dl\la}}
\wt{\cH}_\G=
\frac{d}{dx^\la}(T_\G{}^\la{}_\m+T_M{}^\la{}_\m)+ p^\la_iR^i_{\la\m}
\]
where the term
\[
p_\al{}^{\bt\nu\la} \frac{\dr}{\dr k^\si{}_{\g\dl}}
S^\al{}_{\bt\nu\mu}\frac{\dr}{\dr p_\si{}^{\g\dl\la}}
\wt{\cH}_\G= k^\g{}_{\bt\mu}(\pi_\al{}^{\bt\nu\la}R^\al{}_{\g\nu\la} -
\pi_\g{}^{\al\nu\la}R^\bt{}_{\al\nu\la} -
k^\g{}_{(\bt\mu)}p_\al{}^{\nu\bt\la}R^\al{}_{\nu\g\la}
\]
does not vanish.
In case of gauge theory and gravity without non-metricity, we obtain
\begin{equation}
-\frac1\ve K^\g{}_{(\bt\mu)}R_\al{}^{\nu\bt\la}R^\al{}_{\nu\g\la}
+\frac1{2\kappa} K^\g{}_{(\bt\mu)}R^\bt{}_\la
 - p^\la_iR^i_{\la\m}=
\frac{d}{dx^\la}(T_\G{}^\la{}_\m+T_M{}^\la{}_\m). \label{E20}
\end{equation}
Let us choose the local geodetic coordinate system at a point $x\in X$.
Relative to this coordinate system, the equality (\ref{E20}) at $x$
comes to the conservation law
\[
\frac{d}{dx^\la}(T_\G{}^\la{}_\m+t_M{}^\la{}_\m)=0.
\]

\end{document}